\documentclass[10pt]{elsarticle}
\usepackage{amsmath,amssymb}
\begin{document}

\author{Yamen Hamdouni}
\ead{hamdouniyamen@gmail.com}
\address{Department of Physics, Faculty of Exact Sciences, Mentouri University, Constantine, Algeria}
\title{Random spin distributions and the diffusion equation}
\begin{abstract}
 We show that the probability distribution corresponding to a fully random tracial state of a system of spin-S particles satisfies a diffusion-like equation.
 The diffusion coefficient turns out to be equal
 to  $S(S+1)/6$, where $S$ is the magnitude of the spin of each particle. We also present a bosonization scheme for the lowering and raising total spin operators.
\end{abstract}

 \maketitle
 \section{Introduction}
Many-spin systems exhibit  interesting properties~\cite{kittel}, which explains the reason for which they are  among  the most studied systems in current research. These studies  are motivated 
by several potential applications in mesoscopic and atomic physics, among  which we mention the promising quantum technologies that are slowly emerging. The theoretical
investigation
of the dynamics of spin systems requires, very often, the use of  statistical and probabilistic techniques. In particular, the probability  distribution associated with the 
addition of the spin degrees of freedom  turns out to be  of great importance and usefulness.
For example, we have employed random distributions of spin angular momentum to deal with the description of the decoherence and the entanglement evolution of qubits interacting
with spin environments~\cite{ham1,ham2}. In this letter we address the relation between these Gaussian distributions and the diffusion equation.

To be more explicit in our discussion, we denote by $\vec{J} $ the sum of the  individual spins of  a set of $N$   spin-$S$ particles, i.e., $\vec{J}=\sum_{i=1}^N \vec{S}_i$, and by
$\lambda_j=j(j+1)$ and $m$ the eigenvalues of the operators $\hat J^2$ and $\hat J_z$ respectively. This means that, given a complete set $\{|j,m\rangle\}$ of common eigenvectors of the above operators, we may write
$\hat J^2|j,m\rangle=j(j+1)|j,m\rangle$, and $\hat J_z|j,m\rangle=m |j,m\rangle$. For a given $j$, the multiplicity of $m$ is simple to calculate and is equal to $2j+1$. On the other hand,
the degeneracy $\nu(N,j;S)$ corresponding to the quantum number $j$  is shown to satisfy the relation \cite{ham2}(see the appendix for the proof):
\begin{equation}
 \nu(j,N+1;S)=\sum\limits_{j'=|j-S|}^{j+S}\nu(j',N;S).\label{main}
\end{equation}
 Furthermore, a fully
random tracial state $\omega_N$ of $N$ independent spin-S particles is described by the fully mixed density operator
\begin{equation}
\hat\omega_N= \bigotimes_{i=1}^N \mathbb I_{2S+1}/{ \rm tr}  \mathbb I_{2S+1}=\mathbb I_{2S+1}^{\otimes N}/(2S+1)^N
\end{equation}
where $\mathbb I_{2S+1}$ refers to the $2S+1$-dimensional unit matrix. The action of $ \omega_N$ on any operator $\hat A$ defined on $\mathbb C^{(2S+1)\otimes N}$ (here $\mathbb C$ is
the field of complex numbers) is
prescribed as \cite{vals}
\begin{equation}
 \omega_N(\hat A)={\rm tr}(\hat\omega_N \hat A).
\end{equation}
 With respect to this state,  the quantum number $j$ is assigned a probability, which reads:
\begin{equation}
 P(j,N)=\frac{2j+1}{(2S+1)^N} \nu(j,N;S)\label{rand}.
\end{equation}
The latter is the probability that the eigenvalue of the operator $\hat J^2$ is equal to $j(j+1)$, where, once more, $\vec{J}$ is the sum of the spin operators of $N$
independent spin-S particles. Knowing $P(j,N)$, one can deduce the probability that the eigenvalue of $J_z$ is equal to the quantum number $m$, namely,
\begin{equation}
 P(m,N)=\sum_{j=|m|}^{SN} \frac{P(j,N)}{2j+1}=\sum_{j=|m|}^{SN} \frac{\nu(j,N;S)}{(2S+1)^N},
\end{equation}
which stems from the fact that, for fixed $m$, all the values of $j$ such that $j\geq |m|$ yield the eigenvalue $m$ of the operator $\hat J_z$, and that
for a given $j$, all the values of $m$ such that $-j\leq m \leq j$ are equiprobable.
In what follows, we will be interested in the case where $N$ is sufficiently large. We will show that, under the above condition, equation~(\ref{main})
allows us to prove that the probability distribution $P(j,N)$ 
verifies a diffusion-like equation, that is of great usefulness in deriving its explicit asymptotic Gaussian form.

\section{Explicit form of the probability distribution}
To begin
 we note that it is more convenient to deal with each of the  components of the vector operator  $\vec{J}$ separately; the reasoning could equally  be carried out in terms of vector
 quantities but the notation will be a little cumbersome.
This being said, let $P(m,N)$ be the probability associated with the eigenvalue  $m$ of the operator $\hat J_z$, as defined above. By noting that the multiplicity of $S_z$ is $2S+1$, we infer
from Eq.(\ref{main})
that the following equality holds:
\begin{equation}
 (2S+1)P(m,N+1)=\sum\limits_{\tau=-S}^SP(m-\tau,N).\label{probsum}
\end{equation}
Indeed, we have:
\begin{eqnarray}
 P(m,N+1)&=&\sum_{j=|m|}^{S(N+1)} \frac{\nu(j,N+1;S)}{(2S+1)^{N+1}}\nonumber \\
 &=&\sum_{j=|m|}^{S(N+1)} \frac{1}{(2S+1)^{N+1}} \sum\limits_{j'=|j-S|}^{j+S}\nu(j',N;S).
\end{eqnarray}
It follows that
\begin{equation}
 (2S+1)P(m,N+1)=\sum_{j=|m|}^{S(N+1)}  \sum\limits_{j'=|j-S|}^{j+S} \frac{\nu(j',N;S)}{(2S+1)^{N}}.
\end{equation}
Since $P(m,N)=P(-m,N)$, we assume without loss of generality that $m>0$; then by introducing the variable $\alpha=j-S$, we see that 
\begin{equation}
 (2S+1)P(m,N+1)=\sum_{\alpha=m-S}^{m+S}  \sum\limits_{j'=|\alpha|}^{NS} \frac{\nu(j',N;S)}{(2S+1)^{N}}=\sum_{\alpha=m-S}^{m+S} P(\alpha,N),
\end{equation}
from which equation~(\ref{probsum}) readily follows.

Let us  now focus on the case of $N$ large; in this case, one usually uses Stirling's formula to obtain a Gaussian approximation for $P(m,N)$, which is
fairly good only for values of $m$ that are close to the most probable value of $m$ 
(typically within an interval of the order of the standard deviation of the distribution), which is clearly $m=0$ in our case because of the symmetry.
As we move farther away from this value,  the Gaussian approximation becomes less and less precise, and we should use the exact form of the degeneracy. However,
for very large values of $N$, it is plausible to deal with $N$ and $m$  as  continuous random variables; for instance, $N$ may directly be linked to the density
of particles, while $m$ corresponds to  the $z$-projection of the total classical spin vector of the whole system, which is clearly a continuous quantity. 
Then one is rather dealing with a probability density function, for which
the quantity of interest is
 the probability that $m$ falls within  a certain interval, say $[m,m+dm]$. With that in mind, we may expand in Taylor series, 
 up to second order, both $P(m,N+1)$ and $P(m-\tau,N)$ to obtain

 \begin{equation}
  (2S+1)\biggl[P(m,N)+\frac{\partial P }{\partial N}+\frac{1}{2}\frac{\partial^2 P }{\partial N^2}\biggr]=\sum\limits_{\tau=-S}^S P(m,N)-\frac{\partial P }{\partial m} 
  \sum\limits_{\tau=-S}^S\tau +  \frac{1}{2}\frac{\partial^2 P }{\partial m^2}\sum\limits_{\tau=-S}^S  \tau^2.
 \end{equation}

On account of the fact that
\begin{equation}
 \sum\limits_{\tau=-S}^S  \tau^2=\frac{1}{3} S(2S+1) (S+1),
\end{equation}
we end up with the equation 
\begin{equation}
 \frac{\partial P }{\partial N}+\frac{1}{2}\frac{\partial^2 P }{\partial N^2}=\frac{1}{6}S(S+1)\frac{\partial^2 P }{\partial m^2}.\label{daya}
\end{equation}
From a mathematical point of view, the density $P(m,N)$ is a slowly varying function of $N$. This property  is physically explained by the fact that, when $N$ is very large
(e.g. $N\sim 10^4$), adding or removing a small number of spins  does not affect much the distribution of the system. This means that 
\begin{equation}
 \biggl|\frac{\partial P }{\partial N}\biggr|>> \biggl|\frac{\partial^2 P }{\partial N^2}\biggr| ,
\end{equation}
because $\partial P /\partial N\sim 1/N$ while $\partial^2 P /\partial N^2 \sim 1/N^2$, which allows us to safely neglect the second term in the left-hand side of Eq.(\ref{daya}) when $N$ is large. Whence:
\begin{equation}
 \frac{\partial P }{\partial N}=\frac{1}{6}S(S+1)\frac{\partial^2 P }{\partial m^2}\label{diffspin}.
\end{equation}
It is worth mentioning that the same equation could be derived using a somewhat different method which makes use of the discrete Laplacian. Indeed, we can write 
Eq.(\ref{probsum}) as:
\begin{eqnarray}
 (2S+1)(P(m,N+1)-P(m,N))&=&\sum_{\tau=-S}^{S}(P(m-\tau,N)-P(m,N))\nonumber \\ 
 &\approx&\sum_{\tau=1}^{S} \tau^2 \Delta_\tau P
\end{eqnarray}
where the discrete laplacian is defined by
\begin{equation}
 \Delta_\tau P=\frac{P(m+\tau,N)+P(m-\tau,N)-2P(m,N)}{\tau^2}.
\end{equation}
Now, using the mean value theorem for $\Delta_\tau P$, we arrive at equation (\ref{diffspin}). The latter reminds us the diffusion equation in one dimension without drift, namely,~\cite{zwanz}
\begin{equation}
 \frac{\partial \rho(x,t)}{\partial t}=D\frac{\partial^2 \rho(x,t) }{\partial x^2}, \label{diff}
\end{equation}
where $D$ is the diffusion coefficient. Thus it is tempting  to interpret Eq.(\ref{diffspin}) as being a diffusion equation where the number of spins plays the role of time,
and the role of the position $x$ is played by the magnitude of the spin $m$. The corresponding diffusion coefficient is
\begin{equation}
 D\equiv \frac{1}{6}S(S+1).
\end{equation}

The solution of Eq.(\ref{diffspin}) should be subject to the boundary condition
\begin{equation}
 P(m,0)=0,
\end{equation}
which is physically obvious. The corresponding explicit form can be found by noting that the solution of Eq.(\ref{diff}) is given by
\begin{equation}
 \rho(x,t)=\frac{1}{\sqrt{4\pi D t}}e^{-x^2/(4 D t)}.
\end{equation}
Hence, by comparison, we obtain:
\begin{equation}
 P(m,N)=\sqrt{\frac{3}{2\pi N S(S+1)}}\exp\Bigl\{-\dfrac{3 m^2}{2 N S(S+1)}\Bigr\},\label{dist}
\end{equation}
which is the desired result. 

Some comments would be very useful at this stage. First, in the context of the Brownian motion, we may also write:
\begin{equation}
 P(m',N')=\int_{-\infty}^{+\infty} P(m'-m,N'-N) P(m,N) dN,
\end{equation}
meaning that $P(m'-m,N'-N)$ can be thought of as a transition density, linking the probability densities of the system for different values of the spin magnitudes, and particles
numbers. In the previous expression, however, the semi-group condition $N<N'$ is implied. The latter assertions are equivalent to the well established notion of Greens's function, met in different problems in physics.
Furthermore, notice that the expectation value of $m^2$ turns out to be
\begin{equation}
 \langle m^2\rangle =\frac{NS(S+1)}{3},
\end{equation}
that is 
\begin{equation}
 \langle m^2\rangle =2 D N
\end{equation}
This result is analogous to the Brownian motion mean-square displacement.

Another remark worth mentioning is that  the probability distribution of the random variables associated with the spin operators $\hat J_x/\sqrt{N}$, $\hat J_y/\sqrt{N}$,
 and $\hat J_z/\sqrt{N}$ (which are the components of the operator $\vec{J}=\sum_{i=1}^N\vec{ S}_i$ scaled by $\sqrt{N}$)  when $N\to\infty$ has already  been derived 
 using the trace properties of the angular
 momentum for $S=1/2$~\cite{ham1}.
 Again, the quantity $N$ refers to the total number of the spin-$\frac{1}{2}$
 particles, meaning that for finite $N$, the dimension of the total spin space is $2^N$. For an arbitrary  value of the spin $S$, the same method may be employed to obtain the above
 results as follows: 
 
 The trace of even powers of the components of the total spin operator is given, according to the multinomial theorem, by
 
 \begin{equation}
{\rm tr}\bigl( \hat J_z\bigr)^{2\ell}=\sum_{r_1 r_2\cdots r_N}\frac{2\ell!}{r_1! r_2!\cdots r_N!}{\rm tr} \hat S_{1z }^{r_1}\otimes \hat S_{2z}^{\rm r_2}
\otimes\cdots\otimes \hat S_{Nz}^{r_N}.
\end{equation}
where the natural numbers $r_k$ satisfy $\sum_k r_k=2\ell$. It can be see that the main contribution to the sum comes from the term where all the $r_k$'s are equal to two; then, by
taking into account all the possible partitions of the spins in sets of $\ell$ elements (the remaining terms come from the unit matrix), we find
\begin{eqnarray}
{\rm tr}\bigl(\hat J_z\bigr)^{2\ell}&=&\frac{2\ell!}{\underbrace{2! \times 2!\cdots 2!}_{\ell \ \rm terms }}\sum_{\Pi_\ell} \Biggl(\prod_{k=1}^\ell{\rm tr}\hat S^2_{z k}
\prod_{k=\ell+1}^N{\rm tr}{\mathbb I}_{2S+1}\Biggr)_{\Pi_\ell[1,2,\cdots N]}\nonumber \\&+&Q_{\ell-1}(N),
\end{eqnarray}
 %\end{widetext}
 where as indicated,  the products should be evaluated for all the possible partitions $\Pi_\ell[1,2,\cdots, N]$ of
 $N$ elements into subsets  of $\ell$ elements; the quantity $Q_{\ell-1}(N)$ is a polynomial in $N$ whose degree is at most equal to $\ell-1$. The latter equation means that

 \begin{equation}
  {\rm tr}\bigl(\hat J_z\bigr)^{2\ell}= N^\ell  \Biggl[\frac{(2\ell)!}{2^\ell \ell!} (2S+1)^{N-\ell} [\tfrac{1}{3} S(S+1)(2S+1)]^\ell +O\Biggr(\frac{1}{N}\Biggr)\Biggr].
 \end{equation}

By rescaling the spin operator and taking the limit $N\to\infty$, we obtain

\begin{equation}
 \lim_{N\to\infty}(2S+1)^{-N}{\rm tr}(\hat J_z/\sqrt{N})^{2\ell}=\xi^{2\ell}:=\frac{(2\ell)!}{2^\ell \ell !} [\tfrac{1}{3} S(S+1)]^\ell.\label{mom}
\end{equation}
The characteristic function, with moments given by the right-hand side of Eq.(\ref{mom}), is

\begin{eqnarray}
\Phi(t)&=&\sum_{n=0}^\infty (it)^{2n}\xi^{2n}/(2n)!=\sum_{n=0}^\infty \frac{ (it)^{2n}}{ n! } [\tfrac{1}{6} S(S+1)]^{n}\nonumber \\ &=&\exp\{-\tfrac{1}{6} S(S+1)t^2\},
\end{eqnarray} 
where we used the fact that the odd moments vanish. Afterwards, by Fourier  transforming the characteristic function
\begin{equation}
P(m)=\frac{1}{2\pi}\int_{-\infty}^{\infty}\Phi(t) e^{-it m}dt
\end{equation} 
and making the substitution $m\to m/\sqrt{N}$ we obtain  exactly Eq.(\ref{dist}).

Notice that the joint distribution of three such identical independent random variables, say $m_1,m_2,m_3$,
turns out to be
\begin{eqnarray}
 && P(m_1,m_2,m_3)= P(m_1)P(m_1)P(m_3)\nonumber \\ &&=\Biggl[\frac{3}{2\pi N S(S+1)}\Biggr]^{3/2}\exp\Bigl\{-\dfrac{3 (m_1^2+m_2^2+m_3^2)}{2 N S(S+1)}\Bigr\}.
\end{eqnarray}
Furthermore if we require that $\sqrt{m_1^2+m_2^2+m_3^2}=j$, then the probability distribution becomes~\cite{ham2}
\begin{equation}
\tilde P(m_1,m_2,m_3)\equiv P(j)=4\pi j^2 P(m_1,m_2,m_3).
\end{equation}

\section{Bosonization}

We may now proceed to the bozonization of the scaled lowering and raising operators $\hat J_\pm/\sqrt{N}$. To this end, we stress that given a function
$f(\hat J_x/\sqrt{N}, \hat J_y/\sqrt{N})$ of the spin
operators $\hat J_x/\sqrt{N}$ and $\hat J_y/\sqrt{N}$, we can write:
\begin{equation}
 \lim_{N\to \infty} \omega_N(f(\hat J_x/\sqrt{N}, \hat J_y/\sqrt{N}))=\iint\limits_{-\infty}^{\infty}\bar P(m_1)\bar P(m_2) f(m_1,m_2) dm_1 dm_2,
\end{equation}
where  the probability distribution $\bar P$ is given  (after a proper rescaling) by
\begin{equation}
 \bar P(m)=\sqrt{\frac{3}{2\pi  S(S+1)}}\exp\Bigl\{-\dfrac{3 m^2}{2  S(S+1)}\Bigr\}.
\end{equation}
In fact we can express the above result in terms of the lowering and raising operators, by introducing the polar coordinates  $z$ and $z^*$, with $dz dz^*=dm_1dm_2$, 
such that:
\begin{eqnarray}
 \lim_{N\to \infty} \omega_N(f(\hat J_+/\sqrt{N}, \hat J_-/\sqrt{N}))&=&\frac{3}{2\pi  S(S+1)}\iint_{\mathbb C}f(z^*,z)
 \exp\Bigl\{-\dfrac{3 |z|^2}{2  S(S+1)}\Bigr\} dz^* dz \nonumber \\
 &=&\iint_{\mathbb C}\mathcal P(z^*,z) {\mathcal N}f(z^*,z) dz^* dz.
\end{eqnarray}
We identify the latter equality as being the  coherent state representation of the mean value of $ f $ written in normal order, which is expressed in terms of the so-called 
P-representation~\cite{louis}
\begin{equation}
 \mathcal P(z^*,z)=\sum_n\langle n|\hat\rho\delta(z^*-a^\dag)\delta(z-a)|n\rangle.
\end{equation}
In the above, $a^\dag$ and $a$ are bosonic creation and annihilation  operators, and $\hat \rho$ is a density matrix to be determined. Explicitly we have 
\begin{equation}
 \hat\rho=\iint_{\mathbb C} \mathcal P(z^*,z)|z\rangle \langle z|dz^* dz,
\end{equation}
where $|z\rangle=e^{-|z|^2/2}\sum_{k=0}^{\infty}\frac{z^k}{k!}|k\rangle$ is the usual bosonic coherent state.
Using the polar coordinates $(r,\phi)$,  the elements of the density matrix can be calculated as
\begin{eqnarray}
\langle n|\hat\rho|k\rangle&=&\frac{3}{2\pi  S(S+1)} \int_{z^*} dz^*\int_{z} dz \ \exp\Bigl\{-\dfrac{3 |z|^2}{2  S(S+1)}-|z|^2\Bigr\}
\frac{z^n{z^{*}}^k}{\sqrt{n!k!}}\nonumber \\
&=& \delta_{kn}\frac{3}{  S(S+1)} \int_0^\infty \frac{r^{n+k}}{\sqrt{n!k!}} \exp\Bigl\{-\dfrac{3 r^2}{2  S(S+1)}-r^2\Bigr\} r dr  \nonumber \\
&=&\delta_{kn}\frac{\Bigl(\frac{2S(S+1)}{3}\Bigl)^n}{\Bigl(1+{\frac{2S(S+1)}{3}}\Bigl)^{n+1}}.
\end{eqnarray}
Thus $\hat \rho$ is  the density matrix of a harmonic oscillator which is in thermal equilibrium, at temperature $T$, the value of which
may be obtained by nothing that the expectation value of the occupation 
number $\hat n$ is 
\begin{equation}
 \langle \hat n \rangle=\frac{2S(S+1)}{3}.
\end{equation}
The latter equation immediately yields
\begin{equation}
 \frac{\hbar\omega}{k_B T}=\ln\Bigl(\frac{3+2S(S+1)}{2S(S+1)}\Bigl)
\end{equation}
where $k_B$ is Boltzmann's constant, and $\omega$ is the frequency of the harmonic oscillator.

From the above discussion we conclude that for any well-behaved function of two variables $f$, we can write

\begin{eqnarray}
\lim_{N\to\infty}(2S+1)^{-N}& {\rm tr} & \Biggl\{f\Bigl(\frac{\hat J_+}{\sqrt{N}},\frac{\hat J_-} 
{\sqrt{N}}\Bigr)\Biggr\}=\frac{3}{2S(S+1)}\sqrt{\frac{2S(S+1)}{3+2S(S+1)}}\nonumber \\ &\times &{\rm tr}\Bigl\{e^{-\ln (\frac{3+2S(S+1)}{2S(S+1)})(a^\dag a+\frac{1}{2})}{\mathcal N}f(a^\dag,a)\Bigr\}
\end{eqnarray}
where  $\mathcal N$ denotes the normal ordering. We  thus have constructed a scheme that anables us to map the spin opetarors to their bosonic couterparts. In fact we can also 
 conjecture the following result:
\begin{eqnarray}
\lim_{N\to\infty}(2S+1)^{-N}& {\rm tr} & \Biggl\{ {\mathcal N}f\Bigl(\frac{\hat J_+}{\sqrt{N}},\frac{\hat J_-} 
{\sqrt{N}}\Bigr)\Biggr\}=\frac{3}{2S(S+1)}\sqrt{\frac{2S(S+1)}{3+2S(S+1)}}\nonumber \\ &\times &{\rm tr}\Bigl\{e^{-\ln (\frac{3+2S(S+1)}{2S(S+1)})(a^\dag
a+\frac{1}{2})}f(a^\dag,a)\Bigr\},
\end{eqnarray}
which can be checked by numerical calculation.
\section{Discussion and concluding remarks}
To summarize, we have shown  that the probability distribution corresponding to a random tracial state of a set of $N$ spin-$S$ particles satisfies a 
diffusion-like equation when $N$ is large. We have used this fact to derive its explicit form, and extended the use of the method developed in Ref.~\cite{ham1} to arbitrary
values of the spin. We also were able to come out with a bosonization scheme for the lowering and raising operators. We found that the mean value of the square of the spin is
proportional to the number of particles, the coefficient of proportionality being equal to $2D$.  Hence by knowing the value of $ \langle m^2\rangle$ and $S$
it is possible to deduce the exact value
of $N$, a fact that may exploited experimentally as an indirect method of the measurement of the number of constituents of random spin systems.
Equivalently, if we measurement$ \langle m^2\rangle$ and $N$
it is possible to determine the value of the spin $S$. Moreover, 
the diffusion-like equation would be of great usefulness in studying random spin systems in which transport phenomena are present (e.g by absorptions or adsorption),
when the number of constituents change with time. This would be complementary to ordinary transport equations rendering thus the full investigation of the evolution of
such systems more complete. Another possible application could be in studying high-temperature plasma systems~\cite{hora} where diffusion in space and time is present.
For instance, to verify the validity of the diffusion-like equation, it would be interesting to study a gas of electrons at high temperature~\cite{port}, and to identify the change of
the spin distribution  of the total system as electrons are removed or added to the gas. Finally we note that the use of the explicit form of the distribution makes it possible
to investigate the decoherence of qubits in large  spin systems, which is undoubtedly of great importance. 
\section*{acknowledgements}
The author would like to thank the referees for pointing out the possibility to use the discrete Laplacian,and for the useful comments.
\appendix
\section{}
The aim of this appendix is to derive Eq.(\ref{main})~\cite{ham2}. To this end we note that the spin space of a system of $N$ spin-S particles is given by $\mathbb C^{(2S+1)\otimes N}$.
This space can be decomposed as the direct sum
\begin{equation}
 \mathbb C^{(2S+1)\otimes N}=\bigoplus_{j}^{NS} \nu(j,N;S) \mathbb C^{2j+1}.
\end{equation}
It follows that
\begin{eqnarray}
 \mathbb C^{(2S+1)\otimes (N+1)}&=&\bigoplus_{j}^{(N+1)S} \nu(j,N+1;S) \mathbb C^{2j+1}\nonumber \\
 &=&\bigoplus_{j}^{NS} \nu(j,N;S)   \mathbb C^{2j+1}\otimes \mathbb C^{2S+1} \nonumber \\
 &=&\bigoplus_{j}^{NS} \nu(j,N;S)  \bigoplus_{j'=|j-S|}^{j+S} \mathbb C^{2j'+1}
\end{eqnarray}
By a suitable change of dummy summation indices,  and respecting the triangle rule, we find by comparison that 
\begin{equation}
\nu(j,N+1;S)=\sum\limits_{j'=|j-S|}^{j+S}\nu(j',N;S).
\end{equation}	
%\section*{References}

\end{document}